\documentclass[12pt]{article}
\usepackage{a4wide,epsfig,psfrag,amsmath,amssymb,cite,scalefnt}
\usepackage{color}

\graphicspath{ {figs/} }

\newcommand{\be}{\begin{equation}}
\newcommand{\ee}{\end{equation}}
\newcommand{\bea}{\begin{eqnarray}}
\newcommand{\eea}{\end{eqnarray}}
\newcommand{\ep}{\epsilon}
\newcommand{\Li}[1]{\textrm{Li}_2 \left( #1 \right)}

\allowdisplaybreaks

\newcommand{\gsim}{\;\rlap{\lower 3.5 pt \hbox{$\mathchar \sim$}} \raise 1pt
 \hbox {$>$}\;}
\newcommand{\lsim}{\;\rlap{\lower 3.5 pt \hbox{$\mathchar \sim$}} \raise 1pt
 \hbox {$<$}\;}

\begin{document}

\title{\vskip-3cm{\baselineskip14pt
    \begin{flushleft}
      \normalsize TTP14-024
% LPN14-101
  \end{flushleft}}
  \vskip1.5cm
  Virtual corrections to Higgs boson pair production in the large top quark mass limit
}

\author{
  Jonathan Grigo, 
  Kirill Melnikov\footnote{On leave of absence from
    Department of Physics and Astronomy, Johns Hopkins University, Baltimore,
    MA, USA}, 
  Matthias Steinhauser
  \\[1em]
  {\small\it Institut f{\"u}r Theoretische Teilchenphysik}\\
  {\small\it Karlsruhe Institute of Technology (KIT)}\\
  {\small\it 76128 Karlsruhe, Germany} }

\date{}

\maketitle

\thispagestyle{empty}

\begin{abstract}

  We calculate the three-loop matching coefficient $C_{HH}$, required for a
  consistent description of Higgs boson pair production in gluon fusion
  through  next-to-next-to-leading order  QCD in the heavy top quark
  approximation.  We also compute the 
  $gg \to HH$ amplitude in $m_t \to \infty$ approximation in the full theory
  and show its consistency with an earlier computation in heavy-top effective
  theory.  

%\medskip

%\noindent
%PACS numbers: xxx

\end{abstract}

\thispagestyle{empty}

%- }}}

\newpage

%- {{{ Introduction:

\section{Introduction}

After the discovery of a Higgs boson at the LHC, detailed investigation of its
properties becomes one of primary goals of ATLAS and CMS.  Important among
such studies is the exploration of the Higgs boson self coupling $\lambda$. In
the Standard Model, this coupling is directly related to the Higgs field
potential responsible for the symmetry breaking; in the broken phase, it
induces couplings of three Higgs bosons between themselves.

Experimentally, information about $\lambda$ is obtained from the process of
Higgs boson pair production~\cite{Glover:1987nx,Plehn:1996wb} which will be
accessible after the high-luminosity upgrade of the LHC.  It is
well understood by now that observation of Higgs boson pair production is
difficult and requires both, new ideas on how to isolate the $HH$ signal from
the background, and accurate predictions for the Higgs pair production in the
Standard Model. In the past year we have witnessed significant advances in
both of these directions.

Indeed, building upon the early ideas of Refs.~\cite{Baur:2002qd,Baur:2003gp}
it was suggested to study Higgs pair production in $W^+W^-b \bar b$, $\gamma
\gamma b \bar b$, $b \bar b b \bar b$, and $ b \bar b \tau^+ \tau^-$
channels using substructure
techniques~\cite{Baglio:2012np,Dolan:2012rv,Papaefstathiou:2012qe}, as well as
utilize 
ratios of cross sections~\cite{Goertz:2013eka} for single and double Higgs
production to reduce the theory uncertainty and obtain best sensitivity to
Higgs boson self-couplings.  It remains to be seen how these theoretical ideas
will bare in real experimental searches, but the current consensus seems to be
that the Higgs self-coupling can be measured with the accuracy between twenty
and forty percent (see, e.g., Ref.~\cite{Dawson:2013bba}).

To interpret results of experimental measurements with this accuracy,
one needs to ensure that Standard Model predictions for Higgs boson pair
production are known with sufficient precision. Below we summarize the current
status of theoretical computations of Higgs boson pair production in the
Standard Model.  The leading order predictions for $gg\to HH$ are known since
long ago; they were computed in Refs.~\cite{Glover:1987nx,Plehn:1996wb} where
the exact dependence on all kinematic variables -- primarily the top quark
mass -- has been taken into account.  Improving on these results would have
required the two-loop computations with massive internal (top quarks) and
external (Higgs bosons) particles; currently, such computations are
technically not feasible.  Instead, a possible way forward is provided by
studying the QCD corrections in the approximation where the top quark mass is
taken to be much larger than all other kinematic invariants in the
problem. Working to leading order in $1/m_t$ expansion, one can integrate out
the top quark and obtain an effective theory where Higgs bosons couple
directly to gluons. Within such theory, next-to-leading order (NLO)
computations for 
$pp \to HH$ become feasible and have been performed in
Refs.~\cite{Dawson:1998py} in $m_t \to \infty$ approximation while finite
$1/m_t$ corrections were calculated in~\cite{Grigo:2013rya}.  Recently, the
next-to-next-to-leading order (NNLO)  QCD corrections to $pp \to HH$ were computed
in~\cite{deFlorian:2013uza,deFlorian:2013jea} in $m_t \to \infty$
approximation 
using the close analogy between $pp \to H$ and $pp \to HH$ production in
effective theory.  Soft-gluon resummations and the determination of dominant
$\pi^2$ terms have been considered in~\cite{Shao:2013bz} at
next-to-next-to-leading logarithmic order.

In spite of tremendous progress with fixed order computations for double
Higgs production, we note that NNLO QCD result of
Refs.~\cite{deFlorian:2013uza,deFlorian:2013jea} is formally not complete.
Indeed, at the NNLO QCD accuracy for Higgs pair production, one needs
the Wilson coefficient $C_{HH}$ which was not available when
Refs.~\cite{deFlorian:2013uza,deFlorian:2013jea} were written.  The goal of
this paper is to perform the computation of the $C_{HH}$ Wilson coefficient
and therefore provide the last missing ingredient required to describe the
Higgs boson pair production through NNLO QCD in the large-$m_t$ approximation.

Before we proceed with the computation of the Wilson coefficient, a word of
caution about the validity of large-$m_t$ approximation is in order. Indeed,
it is well-known that for Higgs pair production the $m_t \to \infty$ limit
provides a poor description of both the total cross section and kinematic
distributions. In such a situation it is far from clear that extending $m_t
\to \infty $ computations to NNLO, as was, e.g., done in
Refs.~\cite{deFlorian:2013uza,deFlorian:2013jea}, is a sensible way 
to estimate higher order corrections to Higgs boson pair production. 
Understanding the validity of this approach was the primary goal of
Ref.~\cite{Grigo:2013rya}\footnote{See also
  Refs.~\cite{Grigo:2013xya,Grigo:2014oqa} for higher order terms in the
  expansion in the inverse top quark mass.} where it was shown that, for a
properly chosen leading order cross section, the $1/m_t$ effects at NLO are
moderate, in the $15 - 20$ percent range.  If we assume that the same remains
true at NNLO, we conclude that $m_t \to \infty$ NNLO QCD corrections can be
used to provide a reliable estimate of NNLO QCD corrections with the full top quark mass
dependence.

The remainder of the paper is organized as follows. In the next Section we
introduce the effective Lagrangian for single and double Higgs production in
gluon fusion.  In Section \ref{direct} we describe the matching calculation of
$C_{HH}$. In Section~\ref{sec::chh_2} we discuss the computation of the
virtual corrections to the $gg \to HH$ cross section in the full theory
which serves as the 
cross-check of some results presented in Ref.~\cite{deFlorian:2013uza}. In
Section~\ref{sec::sum} we present our conclusions.

\section{Effective Lagrangian for  Higgs pair production}
\label{effL}

The leading order effective Lagrangian that describes interactions of {\it
  any} number of Higgs bosons with gluons in $m_t \to \infty$ limit is given
by
\begin{equation}
  {\cal L}_{\rm eff} = -\frac{\alpha_s}{3 \pi} {\cal O}_1 \ln \left ( 1 + \frac{H}{v} \right ).
  \label{eq1} 
\end{equation}
In Eq.~(\ref{eq1}) $H$ and $v$ are the Higgs boson field and the vacuum
expectation value, respectively, ${\cal O}_1 = 1/4 G_{\mu\nu}^{ a} G^{\mu
  \nu,a }$, where $G_{\mu\nu}^{ a}$ is the gluon field strength tensor, and
$\alpha_s$ is the strong coupling constant.  This Lagrangian is modified in
higher orders of perturbative QCD.  To account for this, we restrict
Eq.~(\ref{eq1}) to describe  interactions of gluons with up to two Higgs
bosons,\footnote{For the matching coefficients we adopt the notation of
  Ref.~\cite{deFlorian:2013uza}.  This implies that  $C_H\equiv 4C_1$ with $C_1$
  from Ref.~\cite{Chetyrkin:1997un}.} and write
\begin{eqnarray}
  {\cal L}_{\rm eff} &=& - \frac{H}{v} C_H^0 {\cal O}_1^0 +
  \frac{1}{2} \left(\frac{H}{v}\right)^2 C_{HH}^0 {\cal O}_1^0\,, 
  \label{eq::leff}
\end{eqnarray}
The matching coefficients $C_H$ and $C_{HH}$ incorporate radiative effects of
top quarks that are integrated out from the Standard Model; they are given by
perturbative series in the strong coupling constant.

Superscripts ``0'' in Eq.~(\ref{eq::leff}) indicate that 
operator renormalization has not yet been performed, so that both $C_H^0$ and
$C_{HH}^0$ as well as matrix elements involving ${\cal O}_1^0$ are ultraviolet 
divergent. Following Ref.~\cite{Chetyrkin:1997un} we can write $C_X^0 {\cal
  O}_1^0 = C_X^0/Z_{{\cal O}_1} \times Z_{{\cal O}_1} {\cal O}_1^0 = C_X {\cal
  O}_1$, $X \in \{H,HH \}$,  where~\cite{Spiridonov:1984br}
\begin{eqnarray}
  \label{eq::ZO1}
  Z_{{\cal O}_1} &=& 1 
  - \frac{\alpha_s}{4 \pi} \frac{\beta_0}{\epsilon} 
  + \left(\frac{\alpha}{4\pi}\right)^2 
  \left(\frac{\beta_0^2}{\epsilon^2} - \frac{\beta_1}{\epsilon} \right)
  + {\cal O}(\alpha_s^3)
  \,.
\end{eqnarray}
This procedure leads to finite coefficient functions $C_H$ and $C_{HH}$. 
In Eq.~(\ref{eq::ZO1}) we  used $\alpha_s = \alpha_s^{(5)}(\mu)$ to denote the ${\overline {\rm MS}}$ strong 
coupling constant in a theory with five active flavors;  we will use this notation throughout  the paper. 
We also used  standard notation $\beta_0 = 11 C_A / 3 - 4 T_F n_l / 3$ and 
$\beta_1 = 34 C_A^2 / 3 - 4 C_F T_F n_l - 20 C_A T_F n_l / 3$,
where $C_A=N_c, C_F=(N_c^2-1)/(2N_c)$ and $T_F=1/2$ are $SU(N_c)$ color factors
and $n_l=5$ is the number of massless quarks.

It is convenient to introduce the perturbative expansion of $C_H$ and $C_{HH}$ via 
\begin{eqnarray}
\label{eq3}
  C_X &=& - 
  \frac{\alpha_s}{3\pi}\sum_{n\ge0} C_X^{(n)}(\mu)
  \left(\frac{\alpha_s^{(5)}(\mu)}{\pi}\right)^n \,,
  \quad X\in\{H,HH\} \,,
\end{eqnarray}
with $C_H^{(0)}=C_{HH}^{(0)}=1$. Note that equality of $C_H$ and $C_{HH}$ at leading 
order follows from the Lagrangian in Eq.~(\ref{eq1}).  We have chosen to parametrize $C_H$ and
$C_{HH}$ in terms of the  five-flavor strong coupling constant.

%- }}}
%- {{{ Direct calculation of $C_{HH}$:

\section{Direct calculation of matching coefficients}
\label{direct}

Since $C_H$ and $C_{HH}$ are matching coefficients between full and effective theories, it is convenient  to 
derive them as follows: 
   compute amplitudes of  any  physical process 
that depends on one or both of them  in full and effective theories 
and adjust $C_{H}$ and $C_{HH}$ in such a way that the two amplitudes agree. Of course, to determine $C_H$ and $C_{HH}$ independently, 
we need to consider two, rather than one, physical processes;  we choose them  to be {\it i})   Higgs boson production  
in gluon fusion $gg \to H$ and {\it ii}) Higgs boson pair production in gluon fusion $gg \to HH$. The amplitude of the first 
process depends on $C_H$. The amplitude of the second process depends on both $C_H$ and $C_{HH}$. 

We begin with the computation of $C_H$ and  consider the process $g(q_1) g(q_2) \to H$ with $q_1^2 = q_2^2 = 0$ and 
$q_1 \cdot q_2 = m_H^2/2$. We are interested in the behavior of  this process 
in the limit $q_1 \sim q_2 \sim m_H \ll m_t$ where the scattering amplitude can be computed in both full and effective 
theories.   The requirement that the two amplitudes 
are equal up to power-suppressed terms   reads
\begin{equation}
\lim_{q_1,q_2 \to 0} 
 \frac{1}{\zeta_3^{(0)}} {\cal A}^{\rm full}(q_1,q_2,m_H,m_t)  = Z_{{\cal O}_1} {\cal A}^{\rm eff}(q_1,q_2,m_H)+ {\cal O}(q_i/m_t,m_H/m_t).
\label{eq4}
\end{equation}

We now study this equation order-by-order in QCD perturbation theory. At leading order, the amplitude in the full 
theory is given by the one-loop triangle $gg \to H$ diagram which can be
Taylor expanded in external gluon momenta.
The amplitude in the effective theory follows from the Lagrangian Eq.~(\ref{eq1}) 
and reads 
\be
{\cal  A}^{\rm eff} 
= \frac{C_H}{v} \left[ 
  (q_1 \cdot q_2) \epsilon_1 \cdot \epsilon_2 - (\epsilon_1 \cdot q_2)(
  \epsilon_2 \cdot q_1) 
\right]\,. 
\label{eq5}
\ee
Upon equating full and effective theory amplitudes, we find 
$C_H = -\alpha_s/(3 \pi)$, which is the first term in the expansion  
of the result in Eq.~(\ref{eq3}). 

At NLO, the situation changes 
for the following reasons. On one hand, loop corrections to $gg \to H$
amplitudes in the effective theory appear. 
On the other hand, Taylor expansion of $gg \to H$ amplitude in small momenta and the Higgs 
mass  no longer gives correct full theory amplitude even in the limit $q_1 \sim q_2 \sim m_H \ll m_t$ 
since non-analytic dependencies on $s$ and $m_H^2$ do, in general, appear. 

To cure these problems, the large-mass expansion procedure~\cite{Smirnov:2013}
is applied to Feynman diagrams that contribute to the full theory
amplitude. The large-mass expansion splits all loop momenta into soft $k \sim
q_{1} \sim q_2 \sim m_H$ and hard $k \sim m_t$ and allows systematic Taylor
expansions of integrands in both of these regimes.  Scaling of loop momenta
determines scaling of integrals since ${\rm d}^d k|_{\rm soft} \sim s^{d/2}$
and ${\rm d}^d k|_{\rm hard} \sim m_t^{d}$.  Since the $gg \to H$
amplitude necessarily involves at least one loop of top quarks, only one of
the two loop momenta can be soft. For the NLO amplitude in full
theory this implies\footnote{We note that for the process $gg \to H$, $s$ and
  $m_H^2$ are equal.}
\begin{equation}
{\cal A}^{\rm full} = m_t^{-2\ep} {\cal A}_{\rm LO}^{\rm hard} + 
s^{-\ep} m_t^{-2\ep} {\cal A}_{\rm NLO}^{\rm soft} + m_t^{-4\ep} {\cal A}_{\rm NLO}^{\rm hard}. 
\label{eq7}
\end{equation}
We note that {\it hard} part of the amplitude ${\cal A}^{\rm full}$ is
obtained by Taylor expansion of {\it integrands} of loop integrals in powers
of $q_{1,2}/m_t$ and 
$m_H/m_t$; therefore, to obtain ${\cal A}_{\rm NLO}^{\rm hard}$ only two-loop
vacuum integrals need to be computed.  On the contrary, the soft part of the
amplitude requires computation of integrals of form-factor type which depend
on external soft kinematic parameters.  When quantum corrections are computed
in the effective theory, only {\it soft} contributions are
generated. Therefore
\be
{\cal A}^{\rm eff} = C_H \left ( {\cal A}_{\rm LO}^{\rm eff} + s^{-\ep} {\cal A}_{\rm NLO}^{\rm eff} \right ) + ....  
\label{eq8}
\ee
Since we are interested in $C_H$ which, by construction, can not depend on $s$,
Eqs.~(\ref{eq4}),~(\ref{eq7}) and~(\ref{eq8})
can be matched provided that 
\be
 C_H Z_{{\cal O}_1} {\cal A}_{\rm LO}^{\rm eff}  = 
\frac{1}{ \zeta_3^0}
 \left ( m_t^{-2\ep} {\cal A}^{\rm hard}_{\rm LO} + m_t^{-4\ep} {\cal A}_{\rm NLO}^{\rm hard}
\right ).
\label{eq9}
\ee
In Eq.~(\ref{eq9}), 
  $\zeta_3^0$ is  the decoupling
constant of the gluon field (cf. Refs.~\cite{Chetyrkin:1997un,Grozin:2011nk}), which is needed 
for the (on-shell) wave 
function renormalization of external gluons induced by the top quark loops.

The result shown in Eq.~(\ref{eq9}) allows us to obtain the matching coefficient $C_H$ by ignoring all loop corrections 
to $gg \to H$ amplitude in the effective theory {\it and} by computing Taylor expansion 
of relevant diagrams in $q_{1,2}/m_t$ and $m_H/m_t$ in the full theory.   
Extension of the above discussion to  NNLO is straightforward. We write  
\be
 C_H Z_{{\cal O}_1} {\cal A}_{\rm LO}^{\rm eff}  = \frac{1}{ \zeta_3^0}
\left ( m_t^{-2\ep} {\cal A}^{\rm hard}_{\rm LO} + m_t^{-4\ep} {\cal A}_{\rm NLO}^{\rm hard}
+ m_t^{-6\ep} {\cal A}_{\rm NNLO}^{\rm hard}
\right ), 
\label{eq11}
\ee
and solve for $C_H$ order by order in the strong coupling constant $\alpha_s$. 

Before we show the (known) result for $C_H$, we would like to make a few
technical remarks.  First, we note that it may be inconvenient to deal with
external gluon polarization vectors (cf. Eq.~(\ref{eq5})) in multi-loop
computations. If so, one can use an appropriate projection operator to avoid
them. A convenient choice, that respects transversality of the gluon
polarization vectors, is \be \epsilon_1^\mu \epsilon_2^\nu \to -g^{\mu \nu} +
\frac{q_1^\mu q_2^\nu + q_2^\mu q_1^\nu}{q_1 \cdot q_2}, \ee which transforms
the leading order amplitude in Eq.~(\ref{eq5}) into \be {\cal A}^{\rm eff} \to
-\frac{C_H}{v} (d-2) (q_1 \cdot q_2).
\label{eq10}
\ee

Second,  we note that  we first renormalize the top quark mass on-shell, and 
the strong coupling $\alpha_s$
in the $\overline{\rm MS}$
scheme with six active flavors. We then   apply the two-loop
decoupling relations to transform $\alpha_s^{(6)}$
to $\alpha_s^{(5)}$. We note that in this relation the ${\cal O}(\epsilon)$ terms
have to be kept at one-loop order since the two-loop term of $C_H^0$ has an
$1/\epsilon$ pole whereas the one-loop term is finite. The finite result for 
$C_H$, obtained via $C_H^0/Z_{{\cal O}_1}$ is given
by~\cite{Chetyrkin:1997un,Steinhauser:2002rq,Kramer:1996iq}
\be
\begin{split}
  C_H 
  &= -\frac{\alpha_s}{3\pi}\Bigg\{1
  + \left(\frac{5}{4}C_A - \frac{3}{4}C_F \right)\frac{\alpha_s}{\pi} 
  + 
  \Bigg[ 
  \frac{1063}{576}C_A^2 -\frac{5}{96}C_A T_F - \frac{25}{12}C_AC_F 
  + \frac{27}{32}C_F^2 
  \\&
  - \frac{1}{12}C_F T_F
  + \left(\frac{7}{16}C_A^2 - \frac{11}{16}C_AC_F \right) \ln\frac{\mu^2}{m_t^2} 
  + n_l T_F \left ( -\frac{47}{144}C_A - \frac{5}{16}C_F 
    \right.\\&\left.
    + \frac{1}{2}C_F \ln \frac{\mu^2}{m_t^2} \right ) 
  \Bigg]
  \left ( \frac{\alpha_s}{\pi} \right )^2 
  + {\cal O}\left(\alpha_s^3\right)
  \Bigg\}
  % \\&
  \,\,=\, 
  -\frac{\alpha_s}{3\pi}\Bigg\{1
  +\frac{11}{4} \frac{\alpha_s}{\pi} 
  + 
   \left[ \frac{2777}{288} + \frac{19}{16} \ln\frac{\mu^2}{m_t^2}
     \right.\\&\left.
    + n_l \left ( -\frac{67}{96} + \frac{1}{3} \ln \frac{\mu^2}{m_t^2} \right ) 
  \right]
  \left ( \frac{\alpha_s}{\pi} \right )^2 
  + {\cal O}\left(\alpha_s^3\right)
 \Bigg\}\;,
\end{split}
\ee 
where $\alpha_s = \alpha_s^{(5)}(\mu)$ is the ${\overline {\rm MS} }$
coupling constant defined in the theory with $n_l = 5$ massless flavors and
$m_t$ is the pole mass of the top quark.

\begin{figure}[t]
  \begin{center}
    \includegraphics[width=0.7\linewidth]{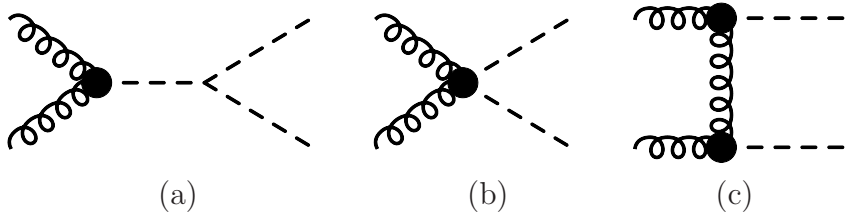}
    \\
    \mbox{}\hspace*{2em}(a) \hspace*{8em} (b) \hspace*{6em} (c)
    \caption[]{\label{fig::gghh_ggh} Effective-theory diagrams with $ggH$ and
      $ggHH$ operators contributing to the double Higgs boson production.}
  \end{center}
\end{figure}

We are now in position to extend the above discussion in such a way that 
  the computation of $C_{HH}$ becomes possible.  To this end, we choose the
gluon fusion process where two Higgs bosons are produced, $g(q_1)g(q_2) \to
H(q_3) H(q_4) $.  We then apply the same reasoning 
as for the single Higgs boson production
and compare amplitudes for ${\cal A}_{gg \to HH}$ computed in full and
effective theories assuming that $q_1 \sim q_2 \sim q_3 \sim q_4 \sim m_H \ll m_t$.  We note,
however, that there is a subtlety in this case that is related to the fact
that pairs of Higgs bosons can not only be produced through the $ggHH$
operator but also through one or two $ggH$ operators in effective theory, see
Fig.~\ref{fig::gghh_ggh}.  This can occur in two different ways.  For example,
already at leading order, double Higgs production in the full theory receives
contributions from a box diagram $gg \to HH$ {\it and} from a triangle
diagram $gg \to H^*$ where the virtual Higgs boson splits into a $HH$ pair.
The second contribution has nothing to do with the matching coefficient
$C_{HH}$.  Our master formula that is based on equating amplitudes in full and
effective theories automatically takes care of this since an identical
contribution is also generated in the effective theory through a local
interaction vertex $ggH$. Hence, diagrams with intermediate off-shell Higgs
bosons cancel {\it exactly} between full and effective theory amplitudes so
that at leading order only $gg \to HH$ box diagram in the full theory is
needed to obtain the Wilson coefficient $C_{HH}$.  Similar subtleties occur in
higher orders, see, e.g., Fig.~\ref{fig::gghh_ggh}(c). Nevertheless, separation
of loop momenta into soft and hard and the understanding that effective theory
loops are always soft allows us to consider only hard contributions in the
full theory and equate them directly to products of matching coefficients and
various tree amplitudes in the effective theory. We therefore obtain the
following generalization of Eq.~(\ref{eq11}) valid in the case of 
Higgs pair production
\be
\begin{split} 
  &   C_{HH} Z_{{\cal O}_1} {\cal A}_{\rm tree, 1PI}^{\rm eff} 
  + C_H^2  Z_{{\cal O}_1}^2 {\cal A}_{\rm tree, 1PR, \lambda = 0}^{\rm eff} 
  + C_H Z_{{\cal O}_1} {\cal A}_{\rm tree, 1PR, \lambda \ne 0 }^{\rm eff}
  \\
  & = \frac{1}{ \zeta_3^0}
  \left (  {\cal A}^{\rm hard}_{\rm 1PI} +  {\cal A}^{\rm hard}_{\rm 1PR, \lambda = 0}
    + {\cal A}^{\rm hard}_{\rm 1PR, \lambda \ne 0}
  \right ).
\end{split} 
\label{eq12}
\ee
When writing Eq.~(\ref{eq12}) we introduced labels ${\rm 1PI}$ and ${\rm 1PR}$, to denote
one-particle reducible and one-particle irreducible contributions in both full
and effective theory. Moreover, we separated various one-particle reducible
contributions on both sides of Eq.~(\ref{eq12}) into those that involve and do
not involve the triple Higgs boson coupling $\lambda$. We also note that these
one-particle reducible contributions contain poles in soft kinematic
parameters, so that it is more appropriate to talk about Laurent rather than
Taylor expansion of full theory amplitudes in Eq.~(\ref{eq12}). However, all
kinematic poles cancel exactly between the left-  and the right-hand side of
Eq.~(\ref{eq12}), as required by the consistency of effective theory.

We note that Eq.~(\ref{eq12}) can be immediately used for the computation of
the matching coefficient $C_{HH}$ since this is the only unknown quantity
there. However, before doing that, it is important to realize that 
Eq.~(\ref{eq12}) can be significantly simplified. Indeed, as the immediate
generalization of the leading order discussion in the previous paragraph, we
observe the exact matching between one-particle reducible contributions to
Eq.~(\ref{eq12}) caused by nonvanishing triple Higgs boson coupling; this
allows us to remove ${\cal A}_{\rm tree, 1PR, \lambda \ne 0 }^{\rm eff}$ and
${\cal A}_{\rm 1PR, \lambda \ne 0 }^{\rm full}$ from both sides of
Eq.~(\ref{eq12}).

It is natural to think that further simplifications are possible. For example,
it is easy to imagine that $Z_{{\cal O}_1}^2C_H^2 {\cal A}_{\rm tree, 1PR, \lambda = 0 }^{\rm eff}$
and ${\cal A}_{\rm 1PR, \lambda = 0 }^{\rm full}$ should match exactly on the
two sides of the equation and can be removed.  Indeed, this is what
happens through two loops but the two contributions do not match exactly at
three loops leaving a remainder that gets re-absorbed into $C_{HH}$ matching
coefficient.  Finally, we want to point out that all calculations have been
performed for arbitrary gauge parameter $\xi$ which drops out in the final
result, a strong check of the correctness of  our calculation.

The final result for $C_{HH}$ that we obtain can be summarized as
follows. Using the parametrization of $C_{H}$ and $C_{HH}$ in Eq.~(\ref{eq3}),
we find 
\be
\begin{split}
  C_{HH}^{(1)} &= C_{H}^{(1)},\;\;\;\; C_{HH}^{(2)} = C_H^{(2)} +  \Delta_{HH}^{(2)}, 
  \\
  \Delta_{HH}^{(2)} &= 
    \frac{7}{8}C_A^2 
  - \frac{5}{6}C_A T_F 
  - \frac{11}{8}C_AC_F 
  + \frac{1}{2}C_F T_F 
  + C_F n_l T_F
  = \frac{35}{24 } + \frac{2 n_l}{3},
  \label{eq:mainr}
\end{split}
\ee
where $n_l$ is the number of massless quarks.  We note that the difference between 
$C_{HH}^{(2)}$ and $C_H^{(2)}$ is significant. Indeed, for $n_l = 5$ and $\mu = m_t$, 
we find 
\be
\Delta_{HH}^{(2)} \approx 4.79,\;\;\; C_H^{(2)} \approx 6.15, 
\label{eq_num}
\ee 
which implies that $C_{HH}^{(2)}/C_H^{(2)} \approx 1.8$.  We note that in
the computation of Refs.~\cite{deFlorian:2013uza,deFlorian:2013jea} it was
assumed that $0 < C_{HH}^{(2)} < 2 C_H^{(2)}$; Eq.~(\ref{eq_num}) shows that
our result for $C^{(2)}_{HH}$ is within this interval but close to its upper boundary.
  The numerical effects
on $C_{HH} \ne C_H$ on the cross section is investigated in Section~\ref{sec::sum}.

%- }}}
%- {{{ Indirect calculation of $C_{HH}$:

\section{\label{sec::chh_2} Virtual corrections to $gg \to HH$ production at NNLO
}

In the previous Section we computed the matching coefficient $C_{HH}$ by
comparing hard contributions in the full theory and tree contributions
in the effective theory. In this way, we only had to compute vacuum bubble
integrals to obtain $C_{HH}$.  However, we can calculate the full $gg \to HH$
amplitude in $m_t \to \infty$ approximation if we account also for soft
contributions in the full theory. Then we obtain the NNLO virtual
corrections to $gg \to HH$ amplitude independent of effective theory
computations.

How difficult is it to compute soft contributions through NNLO for the double Higgs production? It turns out that it is not 
so hard. Indeed, 
since we have to deal with at most  three-loop diagrams in the full theory and 
since at least one of those three loops  has to be hard, the most complicated soft integrals that need to be computed 
are two-loop three-point functions  and 
one-loop four-point functions with all internal and two external lines massless.   All such integrals are known which 
means that we can obtain full $gg \to HH$ amplitude from the full theory.

We consider production of the Higgs boson pair in gluon collisions $g(q_1) + g(q_2) \to H(q_3) + H(q_4)$ and 
introduce Mandelstam variables $s = (q_1+q_2)^2 = (q_3 + q_4)^2$, $ t = (q_1 - q_3)^2 = (q_2 - q_4)^2$ 
and $u = (q_1 - q_4)^2 = (q_2 - q_3)^2$.  Gluons and Higgs bosons are on the mass shell, $q_{1,2}^2 = 0$ 
and $q_{3,4}^2 = m_H^2$.  We write  virtual contributions 
to $gg \to HH$ differential cross section as
\be
\begin{split} 
&  \frac{{\rm d}\sigma_v}{{\rm d}t} 
  =
\frac{{\rm d}\sigma_v^{(0)}}{{\rm d}t}
    + \frac{\alpha_s}{2\pi}                 \frac{{\rm d}\sigma_v^{(1)}}{{\rm d}t}
    + \left(\frac{\alpha_s}{2\pi}\right)^2 \frac{{\rm d}\sigma_v^{(2)}}{{\rm d}t}
    + {\cal O}(\alpha_s^5)
  \,.
\end{split} 
  \label{eq::dsig}
\ee
where again $\alpha_s = \alpha_s^{(5)}(\mu)$.
The leading order cross section in Eq.~(\ref{eq::dsig}) can be written as 
\begin{eqnarray}
  \frac{{\rm d}\sigma_v^{(0)}}{{\rm d}t} &=& 
  \Sigma_{\rm LO} \mathcal{N} \left(C_{\rm LO}^2 - 4 \ep C_{\rm LO}
    + 4 \ep^2\right)
  \,,
\label{eq:lo}
\end{eqnarray}
where
\begin{eqnarray*}
  \mathcal{N} & = & \left(\frac{\mu^2}{m_t^2} \right)^{2 \ep} (1-\ep)
  \left(1 + \ep^2 \zeta_2 - \ep^3 \frac{2}{3} \zeta_3 + \ep^4
    \frac{7}{4} \zeta_4 + \mathcal{O}(\ep^5) \right)
  \,,
\end{eqnarray*}
and\footnote{The definition of $C_{\rm
    LO}$ is taken  from Ref.~\cite{deFlorian:2013uza}, however, we set the
  width of the Higgs boson to zero.}
\be
\begin{split}
  &  \Sigma_{\rm LO} = \frac{ \alpha_s^2  [(tu - m_H^4)/s]^{-\ep} }{
    2^{11} 3^2 v^4 \pi^3 (1-\epsilon)^2 \Gamma(1-\ep) ( 4\pi)^{-\ep}},
  \;\;\;C_{\rm LO} = \frac{6 \lambda v^2}{s-m_H^2} - 1
  \,.
\end{split}
\ee 
We note that $C_{\rm LO}$ is the sum of two leading order contributions to
Higgs boson pair production cross section associated with box and triangle diagrams
and that $\ep$-dependent factors in $\Sigma_{\rm LO}$ originate from the
$d$-dimensional two-particle phase space, and the average over gluon polarizations. 
The higher order $\ep$ terms in Eq.~(\ref{eq:lo}) 
differ from such terms  in Ref.~\cite{deFlorian:2013uza}
since
the matching coefficients used in \cite{deFlorian:2013uza} are strictly four-dimensional. 
We can emulate this effect in our calculation and reproduce the 
results of Ref.~\cite{deFlorian:2013uza}. Similar comments  also apply to the NLO
results given  below.

\begin{figure}[t]
  \begin{center}
    \includegraphics[width=\textwidth]{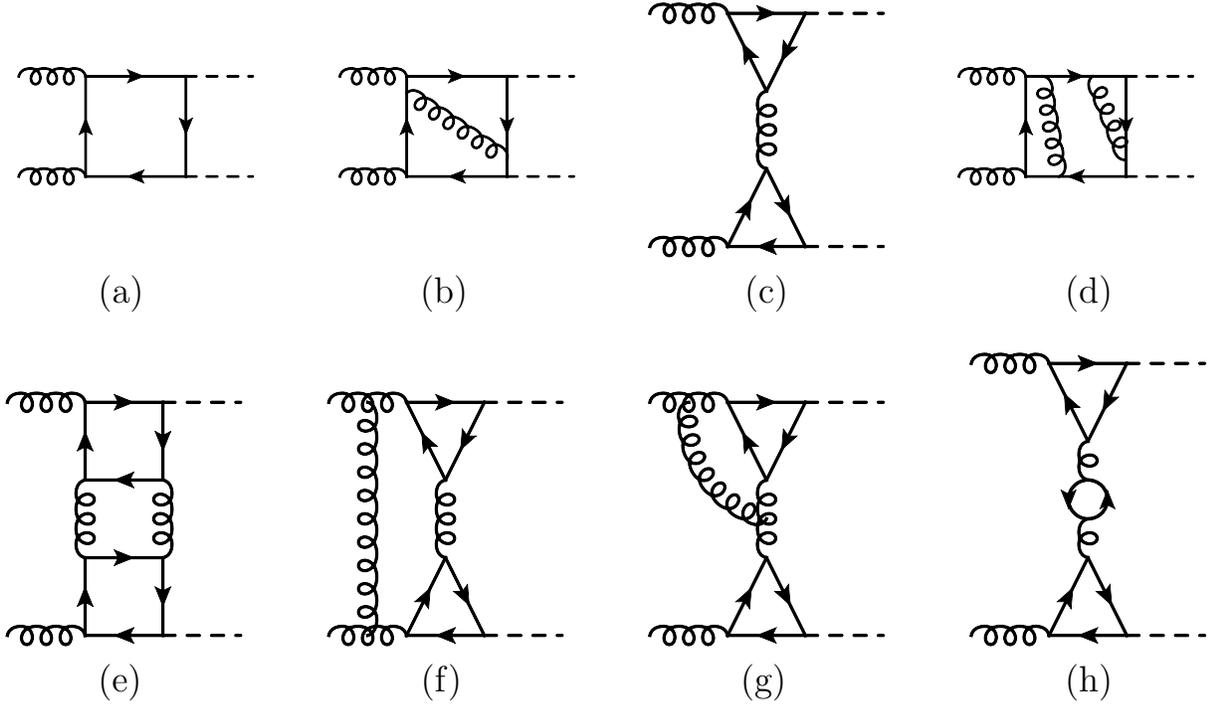}
    \caption[]{\label{fig::m_gghh}
      Sample Feynman diagrams contributing to the amplitude 
      ${\cal A}_{gg\to HH}$.
    }
  \end{center}
\end{figure}

Representative Feynman diagrams contributing to the amplitude ${\cal A}_{gg\to HH}$ can be
found in Fig.~\ref{fig::m_gghh}.  We compute the differential cross sections
using large-mass expansion~\cite{Smirnov:2013} with the help of the {\tt C++}
program {\tt exp}~\cite{q2eexp} that factorizes all integrals into hard
(vacuum) and soft (two-loop three-point and one-loop four-point) integrals.
As we already noticed, all such integrals can be computed in a straightforward
way. Once this is done, we obtain perturbative results  for the virtual corrections
to the  $gg \to HH$ cross section through NNLO in the heavy top approximation.

We note that, since virtual corrections are computed in the full theory, the
results are made ultraviolet finite by means of standard renormalization
procedure. In particular, no matching computations are required. Therefore,
by comparing the result of the full theory computation with
Ref.~\cite{deFlorian:2013uza}, one can independently verify the effective theory
computations reported there and, at the same time, check the consistency of
$C_{HH}$ computation described in the previous Section.

To present the results for virtual corrections, we follow the standard
practice and isolate infrared-divergent pieces using Catani's representation
of scattering amplitudes~\cite{Catani:1998bh}. For ultraviolet finite $gg \to
HH$ scattering amplitude, we write
\be
\begin{split}
  & {\cal A}_{gg \to HH} = \alpha_s \left [ {\cal A}_0 + \frac{\alpha_s}{2 \pi} {\cal A}_1 
    + \left ( \frac{\alpha_s}{2 \pi} \right )^2  {\cal A}_2   \right ],
  \\
  & {\cal A}_1 = I_g^{(1)} {\cal A}_0 + {\cal A}_{1,\rm fin},\;\;\;\;\;
  {\cal A}_2 = I_g^{(2)} {\cal A}_0 + I_g^{(1)} {\cal A}_1 + {\cal A}_{2,\rm fin}.
  \label{eq18}
\end{split}
\ee
The two operators $I_g^{(1,2)}$ depend on QCD color factors $C_{A}$, $C_F$ and $n_l
T_F$, the Mandelstam variable $s$ and the dimensional regularization parameter
$\ep$. In the limit $\ep \to 0$, $I_g^{(1,2)}$ develop $1/\ep^2$ and
$1/\ep^4$ singularities, respectively.  On the other hand, ${\cal A}_{(1,2),\rm
  fin}$ contributions to NLO and NNLO amplitudes are finite.  The exact form
of $I_{g}^{(1,2)}$ operators can be found in
Refs.~\cite{Catani:1998bh,deFlorian:2012za}; we do not reproduce them
here.  Using the representation of scattering amplitude Eq.~(\ref{eq18}),
we write  the virtual contributions to $gg \to HH$ cross sections  as 
\be
\begin{split} 
  & \frac{{\rm d}\sigma_v^{(1)}}{{\rm d}t}
  = \frac{{\rm d}\sigma_{v,\rm fin}^{(1)}}{{\rm d}t}
  + 2 \mbox{Re}\left[I_g^{(1)}\right] \frac{{\rm d}\sigma_v^{(0)}}{{\rm d}t}
  \,,
  \\
  & \frac{{\rm d}\sigma_v^{(2)}}{{\rm d}t}
  = \frac{{\rm d}\sigma_{v,\rm fin}^{(2)}}{{\rm d}t}
  + 2 \mbox{Re}\left[I_g^{(1)}\right] \frac{{\rm d}\sigma_{v,\rm fin}^{(1)}}{{\rm d}t}
  + \left\{ \left|I_g^{(1)}\right|^2 
  + 2 \mbox{Re}\left[\left(I_g^{(1)}\right)^2\right]
  + 2 \mbox{Re}\left[I_g^{(2)}\right]
  \right\}
  \frac{{\rm d}\sigma_v^{(0)}}{{\rm d}t}
  \,,
\end{split} 
  \label{eq::sig_fin}
\ee
It follows from Eq.~(\ref{eq::sig_fin}) that all divergent contributions are proportional 
to either leading or NLO cross sections. Since the leading
order cross section 
has  already been given in Eq.~(\ref{eq:lo}), 
it is sufficient 
to  provide results for finite NLO and NNLO contributions.

\begin{figure}[t]
  \begin{center}
    \includegraphics[width=0.9\linewidth]{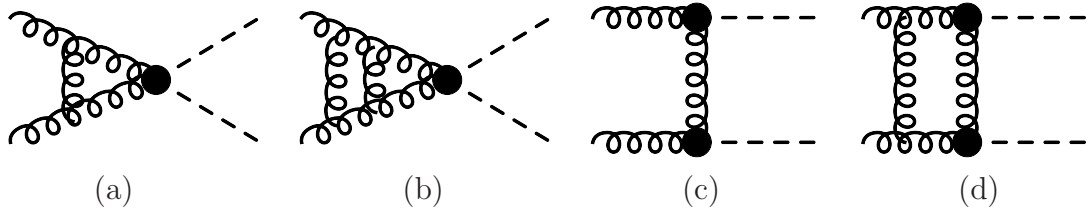}
    \\
    (a) \hspace*{8em} (b) \hspace*{7em} (c) \hspace*{7em} (d) \hspace*{10em}
    \caption[]{\label{fig::gghh_diags}
      One- (a) and two-loop (b) form-factor contributions 
      which lead to ${\cal F}^{(1)}$ and ${\cal F}^{(2)}$.
      Multiplying (c) and (d) with the LO amplitude leads to 
      ${\cal R}^{(1)}$ and ${\cal R}^{(2)}$. ${\cal V}^{(2)}$ is obtained from
      squaring contribution (c).}
  \end{center}
\end{figure}

In the following we present our results in a way which allows for a simple
comparison with Ref.~\cite{deFlorian:2013uza}.  Contributions to $gg\to HH$
amplitude split naturally into two classes -- one that corresponds to only one
effective vertex ($ggH$ or $ggHH$; they occur after shrinking the vacuum
bubbles to a point) and the other one
that involves two $ggH$ vertices. In the former case, all soft contributions
are reducible to three-point functions and are proportional to the leading
order amplitude $C_{\rm LO}$.  Diagrams with two effective vertices start to
contribute at NLO and the corresponding one-loop corrections are needed at NNLO.
For convenience we show sample diagrams up to NNLO in
Fig.~\ref{fig::gghh_diags} where also the notation for the individual
contributions is introduced.  Following this classification, we write the
finite contribution to the one-loop cross section as
\begin{eqnarray}
  \frac{{\rm d}\sigma_{v,\rm fin}^{(1)}}{{\rm d}t}
  &=&
  \Sigma_{\rm LO}\left[
    C_{\rm LO}^2 \left ( \frac{\mu^2}{m_t^2} \right )^{2\ep} {\cal F}^{(1)}
    + C_{\rm LO} \left ( \frac{\mu^2}{m_t^2} \right )^{3\ep} {\cal R}^{(1)}
    \right]
  + {\cal O}\left(\epsilon^3\right)
  \,,
  \label{eq::sig1}
\end{eqnarray}
where the first term in square brackets is the contribution of diagrams with a
single effective vertex and the second term is the contribution of all
diagrams with two effective vertices.  We perform a similar decomposition at
NNLO and write
\begin{eqnarray}
  \frac{{\rm d}\sigma_{v,\rm fin}^{(2)}}{{\rm d}t}
  &=&
  \Sigma_{\rm LO}\left[
    C_{\rm LO}^2 {\cal F}^{(2)}
    + C_{\rm LO} {\cal R}^{(2)}
    + {\cal V}^{(2)}
    \right]
  + {\cal O}\left(\epsilon\right)
  \,,
  \label{eq::sig2}
\end{eqnarray}
where the new element  ${\cal V}^{(2)}$  is the contribution of NLO
diagrams with two effective vertices [cf. Fig.~\ref{fig::gghh_diags}(c)]
 squared.

In addition to soft contribution described so far, hard contributions also
enter equations~(\ref{eq::sig1}) and~(\ref{eq::sig2}). They can be
computed directly using full theory diagrams without resorting to separating
these hard contributions into $C_H$ and $C_{HH}$.  We can then combine $C_{H}$
and $C_{HH}$ results described in the previous Section with the effective 
theory computation reported in Ref.~\cite{deFlorian:2013uza} and compare the
result with the full $m_t \to \infty$ computation described in this
Section.  The two results agree which provides a good consistency check
for both, the effective theory computation and the calculation of the $C_{HH}$
Wilson coefficient reported in the previous Section.

We conclude by showing full results for various quantities that enter
Eqs.~(\ref{eq::sig1}) and~(\ref{eq::sig2}) from the full theory computation.
We give results for arbitrary renormalization scale $\mu$ and separate
contributions due to different color factors.  We obtain

\begin{eqnarray}
  {\cal F}^{(1)} &=& \frac{1}{3} C_A \Big[ 15 + 11 L_s \Big] - 3C_F - \frac{4}{3}
  L_s n_l T_F + \ep \left\{\frac{1}{3} C_A \left[-37 - \frac{77}{2} \zeta_2 + 
     12 \zeta_3 + 15 L_m \right. \right. \nonumber \\
& & \left. \left.  - 11 L_s + \frac{11}{2}L_s^2 \right] + C_F \left[\frac{35}{2} - 3L_m\right] + \frac{1}{3} n_l T_F \Big[14 \zeta_2 + 4 L_s - 2
  L_s^2 \Big] \right\} \nonumber  \\
&&  +  \ep^2 \left\{ \frac{1}{3} C_A \left[\frac{98}{3} +61 \zeta_2
    -\frac{55}{2} \zeta_2 L_s -\frac{47}{3} \zeta_3 + 12 \zeta_3 L_s +
    18 \zeta_4 - 31 L_m + \frac{15}{2} L_m^2  \right. \right. \nonumber \\
&& \left. \left. - 6 L_s - \frac{11}{2}  L_s^2 +
    \frac{11}{6}L_s^3 \right] + \frac{1}{2} C_F \left[-\frac{29}{2} - 9
    \zeta_2 + 35 L_m - 3 L_m^2  \right] 
\right. \nonumber \\
&& \left.  + \frac{2}{3} n_l T_F \left[-7 \zeta_2 + 5 \zeta_2
    L_s  + \frac{2}{3} \zeta_3 + L_s^2 - \frac{1}{3} L_s^3 \right] \right\}
\,, \nonumber \\
{\cal R}^{(1)} &=&  \frac{4}{3} -\frac{2}{3} \ep \left\{1 +\frac{2
      m_H^2}{ s}  + \frac{m_H^4}{tu} -  \frac{2 m_H^6}{ s t u} + C_A
      \Big[45 + 22 L_s \Big] - 45 C_F  -  8 L_s n_l T_F \right\} \nonumber \\
&& + \ep^2 \left\{ 2 \zeta_2 -101 C_F  - \frac{8}{3} n_l T_F \Big[ 7
    \zeta_2 + 2 L_s +2 L_m L_s - L_s^2 \Big] 
\right.  \nonumber \\
&& \left. + \frac{1}{3} C_A \Big[210 + 154 \zeta_2 - 48 \zeta_3 + 22\left(2
    L_s + 2 L_m L_s - L_s^2\right)  \Big] \right\}
  \,,\nonumber\\
   {\cal F}^{(2)} &=&  C_A^2 \left[\frac{23827}{648} - \frac{83}{6} \zeta_2  -
   \frac{253}{36} \zeta_3 + \frac{5}{8} \zeta_4 + \frac{7}{2} L_m +
  \frac{89}{3} L_s + \frac{121}{12} L_s^2 \right] + 9 C_F^2 \nonumber \\
 & &  + C_A C_F \left[-\frac{145}{6} - \frac{11}{2} L_m - 11 L_s
 \right] + n_l^2 T_F^2 \left[ \frac{4}{3} L_s^2 - \frac{22}{9} \zeta_2
 \right]  - \frac{5}{24} C_A - \frac{1}{3} C_F \nonumber \\
 & &  - \frac{1}{3} n_l T_F C_A \left[\frac{2255}{54} + 40 L_s +  22
   L_s^2 - \frac{217}{6} \zeta_2 + \frac{49}{3} \zeta_3 \right] \nonumber \\
 & & -\frac{1}{3}n_l T_F C_F \Big[41 - 12 L_m - 24 \zeta_3 \Big] \,, \nonumber \\
   {\cal R}^{(2)} &=& - 7 C_A^2  + 11 C_A C_F - 8 n_l C_F T_F +
   \frac{1}{3} C_A \left[ \frac{476}{9} + \frac{11}{3} \left(4 L_s + L_t + L_u
     \right) + \frac{4m_H^2}{s}  \right] \nonumber \\
&& - 8 C_F -  \frac{4}{9} T_F n_l  \left[\frac{10}{3} + 4 L_s + L_t +
  L_u \right] -\frac{C_A}{3}\left(1 +
  \frac{2m_H^4}{s^2}\right)\left[2 \Li{1-\frac{m_H^4}{t u}}   \right. \nonumber \\
&& \left.  + 4 \Li{\frac{m_H^2}{t}} + 4
     \Li{\frac{m_H^2}{u}} + 4 \ln\left(1-\frac{m_H^2}{t}\right) \ln
     \left(-\frac{m_H^2}{t}\right) \right. \nonumber \\ 
   & &  \left.  + 
   4 \ln\left(1-\frac{m_H^2}{u}\right) \ln\left(-\frac{m_H^2}{u}\right)
   -8\zeta_2 - \ln^2
  \left(\frac{t}{u}\right)  \right] \,, \nonumber \\
  {\cal V}^{(2)} &=& \frac{1}{(3 s t u)^2}\Big[m_H^8 (t + u)^2 - 2
    m_H^4 t u (t + u)^2 + t^2 u^2 ( 4s^2 + (t + u)^2) \Big]
  \,.
\end{eqnarray}

with $L_m = \ln(\mu^2/m_t^2),\; L_s = \ln(\mu^2/s),\; L_u = \ln[\mu^2/(-u)],
L_t = \ln[\mu^2/(-t)]$.  For $C_A=3$, $C_F=4/3$, $T_F = 1/2$, $\mu^2=s$
    and $\ep = 0$ these results agree with the
analytic expressions of Ref.~\cite{deFlorian:2013uza} provided that
$C_{HH}^{(2)}-C_{H}^{(2)}$ in Eq.~(15) of that reference is replaced by
$\Delta_{HH}^{(2)}$ given in our Eq.~(\ref{eq:mainr}).

%- }}}
%- {{{ Numerical analysis:

\section{Conclusions}
\label{sec::sum}

We computed the three-loop Wilson coefficient  of a $G^2H^2$ operator
that describes interactions of  two Higgs bosons with gluons in the approximation 
that the top quark mass is infinitely large. This is the last missing ingredient that is  required 
to perform consistent NNLO QCD computation of Higgs pair production in the large-$m_t$ limit. 
Our main result -- the three-loop contribution to the Wilson coefficient $C_{HH}$ -- is given 
in Eq.~(\ref{eq:mainr}).  We have also computed virtual corrections to Higgs pair production in gluon 
fusion in the full theory using asymptotic expansions in the inverse top quark mass 
and verified  consistency of our $C_{HH}$ computation with the calculation of $gg \to HH$ 
virtual corrections within  the effective  field theory~\cite{deFlorian:2013uza}. 

An interesting feature of the computed three-loop corrections is that they
break the equality $C_H = C_{HH}$ that persists through two-loops.  Therefore,
their main effect is to change the relative contributions of the box and
triangle diagrams to double Higgs production. Since box and triangle
contributions cancel {\it exactly} at the threshold for producing the two
Higgs bosons, the relatively small difference between $C_H$ and $C_{HH}$ gets
kinematically amplified.\footnote{We note that this is very similar to what
  happens when Higgs boson self-coupling constant $\lambda $ is shifted away
  from its Standard Model value and/or when $1/m_t$ corrections to box and
  triangle contributions are taken into account~\cite{Grigo:2013rya}.}
Indeed, using the relation between Higgs boson self-coupling, the vacuum
expectation value and the Higgs boson mass $2 \lambda^2 v = m_H^2 $, we write
the relative correction as \be \frac{ {\rm d} \sigma_{C_H \ne C_{HH}} - {\rm
    d} \sigma_{C_H = C_{HH}}} {{\rm d} \sigma_{C_H = C_{HH}}} = \frac{2 (
  s-m_H^2)}{(s - 4 m_H^2)} \Delta^{(2)}_{HH} \left ( \frac{\alpha_s}{\pi}
\right )^2 = 0.0117 \left ( \frac{\alpha_s}{0.11} \right )^2 \; \frac{ (
  s-m_H^2)}{(s - 4 m_H^2)}, \ee where $\Delta_{HH}^{(2)}$ from
Eq.~(\ref{eq:mainr}) is used. The strong kinematic enhancement at the
threshold $s = 4 m_H^2$ is evident. Numerically, assuming $m_H = 125~{\rm
  GeV}$ and $\alpha_s = 0.11$, the correction to cross section for $gg \to HH$
computed using $C_H = C_{HH}$ approximation amounts to $6.4$ percent at
$\sqrt{s} = 270~{\rm GeV}$ and $1.7$ percent at $\sqrt{s} = 400~{\rm GeV}$.
The change in the total hadronic cross section $pp \to HH$ amounts to $1\%$,
compared to the case $C_{H}^{(2)}=C_{HH}^{(2)}$.  While all these
corrections are quite moderate, the change in threshold behavior is
interesting and is qualitatively different from a relatively uniform
enhancement of lower-order cross sections provided by soft QCD effects.

%- }}}
%- {{{ Ackn.:

\section*{Acknowledgments}

This work is supported by the Deutsche
Forschungsgemeinschaft through grant STE~945/2-1
and by KIT through its distinguished researcher fellowship program.

%- }}}

%- {{{ bibliography

%- }}}

\end{document}